\documentclass[aps,preprint]{revtex4}

\usepackage{latexsym}

\begin{document}

\preprint{\vbox{\hbox{\tt hep-th/0612158}}}

\title{BFT Embedding of Non-commutative Chiral Bosons}
\author{Wontae Kim}
  \email{wtkim@sogang.ac.kr}
  \affiliation{Department of Physics and Center for Quantum Spacetime,
    Sogang University, C.P.O. Box 1142, Seoul 100-611, Korea}
\author{Young-Jai Park}
  \email{yjpark@sogang.ac.kr}
  \affiliation{Department of Physics and Center for Quantum Spacetime,
    Sogang University, C.P.O. Box 1142, Seoul 100-611, Korea}
\author{Hyeonjoon Shin}
  \email{hshin@sogang.ac.kr}
  \affiliation{Center for Quantum Spacetime, Sogang University, Seoul
    121-742, Korea} 
\author{Myung Seok Yoon}
  \email{younms@sogang.ac.kr}
  \affiliation{Center for Quantum Spacetime, Sogang
    University, Seoul 121-742, Korea}

\date{\today}

\begin{abstract}
A two dimensional model of chiral bosons in non-commutative field space
is considered in the framework of the Batalin-Fradkin-Tyutin (BFT) Hamiltonian
embedding method converting the second-class constrained system into
the first-class one.  The symmetry structure associated with the first-class
constraints is explored and the propagation speed of fields is
equivalent to that of the second-class constraint system.
\end{abstract}


\maketitle

\section{Introduction}
\label{sec:intro}

The model of chiral bosons in two dimensions is basically a constrained 
system.  Although it is simple, the study of its structure may give us
helpful insights in understanding various models with some chiral
structure, like theories of self-dual objects appearing in superstring 
theory.  One of the major issues about chiral boson is how to treat its 
constraint structure consistently in the canonical Hamiltonian formalism.  
Due to the interesting feature of the model itself and its implicit relevance 
to other models, there have been lots of studies from the various viewpoints,
which are based on largely three approaches,
the Floreanini-Jackiw (FJ) method \cite{fj}, 
the method with linear constraint \cite{sriv}, and
the Batalin-Fradkin-Tyutin (BFT) embedding method \cite{bft}.

The usual playground for the study of chiral bosons is assumed to be 
restricted to the commutative field space.   
On the other hand, there is some arguments about the possibility that
some non-commutative effects could take place in ultra high-energy
physics without violating the Lorentz invariance. Motivated by this, one may
consider the chiral bosons in the non-commutative field space, based on
achievements of the previous studies.  Indeed, the construction of the
corresponding model has been given, and the problems on the bosonization
and the Lorentz invariance have been studied in Ref.~\cite{dgml}.

Having the model of chiral bosons in the non-commutative field space,
it is natural to ask about its canonical structure and investigate, if
any, its difference from the commutative model. 
In this Brief Report,
we study the canonical structure of the model in the framework of the
BFT embedding method \cite{bft}.  The BFT method converts the
second-class constrained system into the first-class one by
introducing auxiliary fields and hence extending the phase space, and
allows one to have local symmetries associated with the first-class
constraints (See Refs.~\cite{mwy,wot,an,amw} for chiral bosons,
Refs.~\cite{ban,kp,brr,bcs,bnon,kkp,kwp} for Chern-Simons
model, and Ref.~\cite{hkpy} for non-commutative D-brane system).  The
resulting full first-class 
constrained system in the extended phase space usually has many fields
(infinite number of fields in our case).  We consider the propagation
speed of each field and investigate the consistency in Lorentz invariance.

The organization of this paper is as follows.
In the next section, we introduce the model of chiral boson in 
non-commutative field space and take into account of its second-class
constraints via the method of symplectic structure \cite{fj}. 
The BFT embedding of the model
follows in Sec.~\ref{sec:bft} and the second-class constraints are fully
converted into the first-class one in the extended phase space.  The
resulting extended system is shown to have infinite local symmetries.  
At the end, from the equations
of motion of the fields, the propagation speed in the non-commutative
field space is considered. Conclusions are given in Sec.~\ref{sec:dis}.


\section{Non-commutative Chiral Boson}
\label{sec:com}

The non-commutativity in field space is basically represented by the 
non-vanishing commutator between different elementary fields.
The action for a theory in non-commutative field space is constructed in a
way that such non-commutativity is realized.
The theory of chiral boson in non-commutative field space has been
constructed in Ref.~\cite{dgml}. In order to study the theory of a
chiral boson in a non-commutative field space, the Poisson brackets in
this model have been deformed by the non-commutative
parameter $\theta$. In this case, the action is given by
\begin{equation}
  \label{action-0}
  S = \int d^2 x \left[ - \frac{2}{1+\theta^2} {\dot\phi}_a \Delta_{ab}
  \phi'_b - \phi'_a \phi'_a \right],
\end{equation}
where the overdot and the prime denote the derivatives with respect to time
and space, respectively \cite{dgml}.
The left (right) moving field is represented by the subscript
$a$ with positive (negative) sign.
The $2\times 2$ matrix $\Delta_{ab}$ encodes the non-commutativity
of the field space with the non-commutative parameter $\theta$ and
is defined by
\begin{eqnarray}
  \Delta_{ab} &\equiv& \frac{a}{2}(\theta\epsilon_{ab} -\delta_{ab})
       \nonumber \\
    &=& \frac12 \left( 
      \begin{array}{cc}
        -1 & \theta \\
        \theta & 1
      \end{array}
      \right)~,  \label{Delta}
\end{eqnarray}
the inverse of which is
\begin{equation}
  \label{inverseD}
  \Delta_{ab}^{-1} = \frac{4}{1+\theta^2} \Delta_{ab} ~.
\end{equation}
Note that for the $\theta\to 0$ limit, the action reduces to the FJ
action \cite{fj}.

The system (\ref{action-0}) is basically a constrained one,
since the canonical momentum $\Pi_a$ of the field $\phi_a$ does not contain
any time evolution of the field as can be easily seen by
\begin{equation}
  \label{mom-0}
  \Pi_a = - \frac{2}{1+\theta^2} \Delta_{ab} \phi'_b  ~.
\end{equation}
The primary constraints are then
\begin{equation}
  \label{constr}
  \Omega_a = \Pi_a + \frac{2}{1+\theta^2} \Delta_{ab} \phi'_b \approx 0~,
\end{equation}
and the evaluation of the Poisson bracket between them gives
\begin{equation}
  \label{C}
\{ \Omega_a(x), \Omega_b(y) \} = \frac{4}{1+\theta^2} \Delta_{ab}
  \partial_x \delta(x-y) ~,
\end{equation}
from which we see that the primary constraints are in the second-class.
The time evolution of primary constraints by using the primary Hamiltonian
defined by 
$H_p = H_c + \int dx \lambda_a \Omega_a$,
where $H_c$ is the canonical Hamiltonian corresponding to the
action (\ref{action-0}),
\begin{equation}
  \label{H-0}
  H_c  = \int dx \phi'_a \phi'_a ~,
\end{equation}
results in fixing the Lagrangian multiplier fields $\lambda_a$.
Therefore, the primary constraints (\ref{constr}) form a full set of 
constraints for the system (\ref{action-0}).

All the constraints are in second-class and hence a proper procedure
is required to implement them consistently.  Although the usual Dirac
procedure \cite{dir} may be considered, we take a more smart method
based on the symplectic structure developed by Floreanini and Jackiw
(FJ) \cite{fj}, which is basically concerned with symplectic
structure. Since our system (\ref{action-0}) is just the first order
one, the symplectic structure method is especially suitable. The
symplectic structure, say $C_{ab}$, is read off from the first order
term in time derivative, and its precise form in the present case is
obtained as $C_{ab}=\frac{4}{1+\theta^2} \Delta_{ab} \partial_x
\delta(x-y)$. The so called FJ bracket between the field variables
$\phi_a$  is simply given by the inverse of the symplectic structure,
which is 
\begin{equation}
  C^{-1}_{ab}(x,y) = \Delta_{ab} \frac{1}{\partial_x} \delta(x-y) =
  \Delta_{ab} \epsilon(x-y) ~,
\end{equation}
where $\epsilon(x-y)$ is the step function.  The resulting non-vanishing
brackets between elementary fields are then obtained as follows.
\begin{eqnarray}
  \{ \phi_a(x), \phi_b(y) \}_{\rm FJ} &=& \Delta_{ab} \epsilon(x-y) ~, \\
  \{ \phi_a(x), \Pi_b(y) \}_{\rm FJ} &=& \frac12 \delta_{ab} \delta(x-y) ~,
  \\
  \{ \Pi_a(x), \Pi_b(y) \}_{\rm FJ} &=& - \frac{1}{1+\theta^2}
  \Delta_{ab} \partial_x \delta(x-y) ~.
\end{eqnarray}
There are equivalent to the Dirac brackets and they recovers the
conventional brackets for chiral bosons for the $\theta \to 0$ limit.


\section{BFT Embedding}
\label{sec:bft}

In this section, we consider the system (\ref{action-0}) in the
framework of the BFT Hamiltonian embedding method and converts it into
the first-class constrained system. For notational convenience, we
replace the fields $\phi_a$ and $\Pi_a$ with $\phi_a^{(0)}$ and
$\Pi_a^{(0)}$ respectively. The second-class constraints
(\ref{constr}) are then written as
\begin{equation}
  \label{constr-0}
  \Omega_a^{(0)} = \Pi_a^{(0)} + \frac{2}{1+\theta^2} \Delta_{ab}
  \phi^{(0)\prime}_b \approx 0 ~. 
\end{equation}
In order to convert these constraints into first-class one, we first
extend the phase space by introducing auxiliary fields $\phi_a^{(1)}$ 
(one auxiliary field for each constraint), which satisfy
\begin{equation}
  \label{phi-1}
  \{ \phi_a^{(1)}(x), \phi_b^{(1)}(y) \} = \gamma_{ab}(x,y) ~,
\end{equation}
with $\gamma_{ab}$ determined later on.

In the extended phase space, a proper modification of constraints 
$\Omega_a^{(0)}$ is given by 
\begin{equation}
  \label{constr-1}
  \tilde\Omega_a^{(0)} = \Omega_a^{(0)} 
  + \sum_{k=1}^\infty \omega_a^{(1,k)},
\end{equation}
which have to satisfy the boundary condition 
$\tilde\Omega_a^{(0)} |_{\phi_a^{(1)}=0} = \Omega_a^{(0)}$ and the 
requirement of strong involution, 
$\{ \tilde\Omega_a^{(0)}, \tilde\Omega_b^{(0)} \} =0$,
to accomplish the BFT embedding.  Here we would like to note that
the strong involution is valid only for the Abelian theory, which is the
case at hand.  As for the non-Abelian case, the weak involution should
be considered.
The correction $\omega_a^{(1,k)}$ at
a given order $k$ is assumed to be proportional to $(\phi_a^{(1)})^k$.
To begin with,  we consider the first order correction which is given by
\begin{equation}
  \label{Omega0-corr}
\omega_a^{(1,1)} = \int dy \, X_{ab}(x,y) \phi_b^{(1)}. 
\end{equation}
It is not so difficult to show that the requirement of strong involution 
leads us to have the simple solution
for $\gamma_{ab}$ of Eq.~(\ref{phi-1}) and $X_{ab}$ as
\begin{eqnarray}
  \gamma_{ab}(x,y) &=& \Delta_{ab} \epsilon(x-y) ~, 
               \label{gamma} \\
  X_{ab}(x,y) &=& \frac{4}{1+\theta^2} \Delta_{ab} \partial_x
  \delta(x-y) ~. 
             \label{X}
\end{eqnarray}
We see that the constraints (\ref{constr-1}) become the first-class one 
already at the level of the first correction.  This means that it is
not necessary to consider higher order corrections and hence we can safely
set them to zero. 
The resulting first-class constraints in the phase space extended
by introducing the fields $\phi_a^{(1)}$ is then
\begin{equation}
  \label{Omega0}
  \tilde\Omega_a^{(0)} = \Pi_a^{(0)} + \frac{2}{1+\theta^2}
  \Delta_{ab} (\phi_b^{(0)})' + \frac{4}{1+\theta^2} 
  \Delta_{ab} (\phi_b^{(1)})'  ~.
\end{equation}

The canonical Hamiltonian $H_c^{(0)} \equiv H_c$ of Eq.~(\ref{H-0})
is the one only for the fields $\phi^{(0)}_a$.  Similar to the modification
of constraints in Eq.~(\ref{constr-1}), it should also be modified 
properly in the extended phase space.  The new canonical Hamiltonian
is defined by
$H_c^{(1)} = H_c^{(0)} + h^{(1)}$, where $h^{(1)}$ is determined from 
the involutive condition 
$\{ \tilde\Omega_a^{(0)} , H_c^{(1)} \} = 0$.  In the present case,
what we get is 
\begin{equation}
  \label{H1}
H_c^{(1)} = \int dx \left( (\phi_a^{(0)})' +
  (\phi_a^{(1)})' \right) \left( (\phi_a^{(0)})' +
  (\phi_a^{(1)})' \right).
\end{equation}
Given this Hamiltonian, we can obtain the corresponding Lagrangian by
considering the partition function to explore the constraint structure
in the extended phase space.  The phase space partition function is 
given by
\begin{equation}
  \label{Z}
  Z = \int \prod_{a=\pm} \prod_{n=0,1} {\cal D}\phi_a^{(n)} {\cal D} \Pi_a^{(0)} \delta[\tilde
  \Omega_a^{(0)}] \delta[\Gamma_a^{(0)}] \det|\{\tilde\Omega_a^{(0)} ,
  \Gamma_a^{(0)}\}| \ e^{iS^{(1)}},
\end{equation}
where 
\begin{equation}
  \label{S1}
  S^{(1)} = \int d^2 x \left( \Pi_a^{(0)} {\dot\phi}_a^{(0)} +
  \frac12 \int dy \phi_a^{(1)}(x) \gamma_{ab}^{-1}(x,y)
  {\dot\phi}_b^{(1)}(y) \right) - \int dt H_c^{(1)},
\end{equation}
and $\Gamma_a^{(0)}$ are gauge fixing conditions to make the non-vanishing
determinant of $\tilde \Omega_a^{(0)}$ and $\Gamma_a^{(0)}$.
Through the usual procedure of path integration with respect to 
the momenta $\Pi_a^{(0)}$ and by noticing
from Eq.~(\ref{gamma})
\begin{equation}
  \gamma_{ab}^{-1} (x,y) = \frac{4}{1+\theta^2} \Delta_{ab}
     \partial_x \delta(x-y),
\end{equation}
the Lagrangian density ($S^{(1)} = \int d^2x\, \mathcal{L}^{(1)}$)
is obtained as
\begin{eqnarray}
  {\cal L}^{(1)} &=& - \frac{2}{1+\theta^2} \left( {\dot\phi}_a^{(0)}
    \Delta_{ab} (\phi_b^{(0)})' + {\dot\phi}_a^{(1)}
    \Delta_{ab} (\phi_b^{(1)})' \right) - (\phi_a^{(0)})'(\phi_a^{(0)})'
    - (\phi_a^{(1)})'(\phi_a^{(1)})' \nonumber \\ 
  & & - \frac{4}{1+\theta^2} {\dot\phi}_a^{(0)}
    \Delta_{ab} (\phi_b^{(1)})' - 2 (\phi_a^{(0)})'(\phi_a^{(1)})'.
    \label{L0}
\end{eqnarray}
From this Lagrangian, the canonical momenta conjugate to
$\phi_a^{(0)}$ and $\phi_a^{(1)}$ are derived as 
\begin{eqnarray}
  \Pi_a^{(0)} &=& - \frac{2}{1+\theta^2} \Delta_{ab} (\phi_b^{(0)})' -
     \frac{4}{1+\theta^2} \Delta_{ab} (\phi_b^{(1)})' ~, \\
  \Pi_a^{(1)} &=& - \frac{2}{1+\theta^2} \Delta_{ab} (\phi_b^{(1)})' ~,
\end{eqnarray}
which lead to the following constraints:
\begin{eqnarray}
  \tilde\Omega_a^{(0)} &=& \Pi_a^{(0)} + \frac{2}{1+\theta^2} \Delta_{ab}
  (\phi_b^{(0)})' - \left( \Pi_a^{(1)} - \frac{2}{1+\theta^2} \Delta_{ab}
  (\phi_b^{(1)})' \right) \approx 0 ~,  \label{c0} \\
  \Omega_a^{(1)} &=& \Pi_a^{(1)} + \frac{2}{1+\theta^2} \Delta_{ab}
  (\phi_b^{(1)})' \approx 0 ~, \label{c1}
\end{eqnarray}
where the constraints $\tilde \Omega_a^{(0)}$ has been rewritten
by using the constraints $\Omega_a^{(1)}$.  The time evolution of these
constraints via the primary Hamiltonian based on $H_c^{(1)}$ gives 
no more constraints and thus we see that 
$\tilde \Omega_a^{(0)}$ and $\Omega_a^{(1)}$ form a full set of 
constraints for the system described by $\mathcal{L}^{(1)}$.

The Poisson bracket structure between the constraints, Eqs.~(\ref{c0})
and (\ref{c1}), shows that $\Omega_a^{(1)}$ are in second-class, while
$\tilde \Omega_a^{(0)}$ are the first-class constraints as expected.
(Throughout this work, the first-class constraints are denoted with
tilde.  This is why we do not put tilde on $\Omega_a^{(1)}$.)  
This means that the system in the extended phase space is not a fully 
first-class constrained one and the procedure of BFT embedding is not yet
completed.  At this point, we observe that $\Omega_a^{(1)}$ is exactly 
the same as $\Omega_a^{(0)}$ in Eq.~(\ref{constr-0}) if $\phi_a^{(0)}$
and $\Pi_a^{(0)}$ are substituted for $\phi_a^{(1)}$ and $\Pi_a^{(1)}$
respectively.  By introducing another auxiliary fields, say $\phi_a^{(2)}$, 
and taking the same steps from Eq.~(\ref{constr-0}) to Eq.~(\ref{c1}),
we can convert $\Omega_a^{(1)}$ into the first-class constraints
$\tilde \Omega_a^{(1)}$.  However, the canonical momenta $\Pi_a^{(2)}$
of $\phi_a^{(2)}$ give new constraints $\Omega_a^{(2)}$ which are in
second-class.  It is necessary to introduce the third auxiliary fields 
$\phi_a^{(3)}$, and the story continues forever.  As a result, the present situation
requires the introduction of infinitely many auxiliary fields to accomplish 
the BFT embedding procedure.  This in turn implies that the extended phase 
space is of infinite dimensionality.  We note that this kind of infinite 
dimensional extended phase space appears also in the study of Abelian
Chern-Simons theory \cite{kkp}.

Then, the infinite repeat of the BFT embedding method gives us finally the
canonical Hamiltonian of the fully first-class constrained system, which is
\begin{equation}
  \tilde H_c = \int dx \sum_{n=0}^\infty \sum_{m=0}^\infty (\phi_a^{(m)})'
    (\phi_a^{(n)})' ~. \label{H:final}
\end{equation}
The corresponding Lagrangian is obtained as 
\begin{eqnarray}
  {\cal L} &=& \sum_{n=0}^\infty \left[ -\frac{2}{1+\theta^2}
    {\dot\phi}_a^{(n)} \Delta_{ab} (\phi_b^{(n)})' -
    (\phi_a^{(n)})'(\phi_a^{(n)})' \right] \nonumber \\
    & & +2 \sum_{n=1}^\infty \sum_{m=0}^{n-1} \left[ -\frac{2}{1+\theta^2}
    {\dot\phi}_a^{(m)} \Delta_{ab} (\phi_b^{(n)})' -
    (\phi_a^{(m)})'(\phi_a^{(n)})' \right] ~. \label{L:final}
\end{eqnarray}
From the canonical momenta $\Pi_a^{(n)}$ conjugate to the fields
$\phi_a^{(n)}$,
\begin{equation}
  \label{mom-n}
  \Pi_a^{(n)} = - \frac{2}{1+\theta^2} \Delta_{ab} (\phi_b^{(n)})' 
  - \frac{4}{1+\theta^2} \Delta_{ab} \sum_{m=n+1}^\infty (\phi_b^{(m)})' ,
\end{equation}
we get the constraints
\begin{equation}
  \label{Omega-n}
\tilde\Omega_a^{(n)} = \Pi_a^{(n)} + \frac{2}{1+\theta^2} \Delta_{ab}
  (\phi_b^{(n)})' - \left( \Pi_a^{(n+1)} - \frac{2}{1+\theta^2} \Delta_{ab}
  (\phi_b^{(n+1)})' \right) \approx 0 ~,
\end{equation}
which are in first-class, as it should be. It can be easily checked
that the constraints (\ref{Omega-n}) satisfies 
$\{ \tilde\Omega_a^{(n)}(x), \tilde H_c \} = 0$. 

Now we are in a position to be able to investigate new local symmetries 
of the first-class constrained system (\ref{L:final}). The total action is 
written as
\begin{equation}
  \label{S:final}
  S = \int d^2 x \sum_{n=0}^\infty \Pi_a^{(n)} \dot\phi_a^{(n)} - \int
  dt \tilde H_c + \int d^2 x 
   \sum_{n=0}^\infty \lambda_a^{(n)} \tilde\Omega_a^{(n)},
\end{equation}
where $\lambda_a^{(n)}$'s are Lagrange multipliers. 
It can be shown that the action is 
invariant under the following local gauge transformations:
\begin{eqnarray}
  \delta \phi_a^{(n)} &=& - \epsilon_a^{(n)} + \epsilon_a^{(n-1)},
     \label{dphi} \\
  \delta \Pi_a^{(n)} &=& - \frac{2}{1+\theta^2} \Delta_{ab}
     [(\epsilon_a^{(n)})' + (\epsilon_a^{(n-1)})'], \label{dPi} \\
  \delta \lambda_a^{(n)} &=& -\dot\epsilon_a^{(n)}, \label{dlambda}
\end{eqnarray}
where $\epsilon_a^{(n)}(x)$ are infinitesimal gauge parameters with
$\epsilon_a^{(-1)}=0$ and $n$ is non-negative integer valued.
As is well established, these local symmetries are generated by
the first-class constraints.  Since there are infinite number of 
first-class constraints in the present situation, the model we are
considering has infinite local symmetries.

Finally, we consider the propagation of fields in the non-commutative
field space.  The equations of motion for the fields $\phi_a^{(n)}$
are derived from the variation of the Lagrangian (\ref{L:final}) as
\begin{equation}
  \label{eom}
  \sum_{n=0}^\infty [ \dot\phi_a^{(n)} 
  + 2 \Delta_{ab} (\phi_b^{(n)})' ]=0 ~.
\end{equation}
where Eq.~(\ref{inverseD}) has been used.
In light-cone coordinates $x^\pm$ ($\equiv (ct \pm x)/2$), these
equations split into two parts
\begin{eqnarray}
  \sum_{n=0}^\infty \partial_- \phi_+^{(n)} &=& - \theta
    \sum_{n=0}^\infty (\phi_-^{(n)})' , \label{eom:1} \\ 
  \sum_{n=0}^\infty \partial_+ \phi_-^{(n)} &=& - \theta
    \sum_{n=0}^\infty (\phi_+^{(n)})' ~, \label{eom:2} 
\end{eqnarray}
from which we can obtain 
\begin{equation}
  \label{eom:final}
  \sum_{n=0}^\infty (\Box - \theta^2 \partial_x^2 ) \phi_a^{(n)} = 0 ~,
\end{equation}
where $\Box \equiv (1/c^2)\partial_t^2 - \partial_x^2$.  This means that,
by the effect of the non-commutativity in the field space,
the propagation speed of the fields is modified to
\begin{equation}
  \label{light}
  c \to c'=c\sqrt{1+\theta^2} ~,
\end{equation}
which was noticed in Ref.~\cite{dgml}.  As was pointed
out by the authors of \cite{dgml}, however, this modification of the 
propagation speed does not mean the violation of Lorentz invariance.
The present formulation in the framework of the BFT embedding 
method shows that such modification takes place for all the
fields with exactly the same manner, and thus does not lead to any 
inconsistency in Lorentz invariance.  



\section{Conclusion}
\label{sec:dis}
We have shown that the second class constraint system for the chiral
bosons in the non-commutative field space has been converted into the
first class constraint system by using the BFT method, where the
resulting brackets can be implemented by the conventional Poisson
algebra. The resulting equation of motion
(\ref{eom:final}) is symmetric under the transformation of
(\ref{dphi}), which can be shown by the total summation of the
infinitesimal transformation parameters are canceled completely. In
general, the original second class constraint system can be
interpreted as a gauge fixed version of the first class constraint
system in the context of the BFT method. Therefore, the equation of
motion (\ref{eom:final}) and (\ref{light}) have been derived in a
gauge independent fashion. Of course, each scalar field in the first
class constraint system has the same velocity with that of
the velocity in the gauge fixed system corresponding to the second
class constraint system if the auxiliary field $\phi^{(n)}_a$ has its
angular frequency $w^{(n)}_a$ and the wave number $k^{(n)}_a$,
respectively.


\section*{Acknowledgments}
This work was supported by the Science Research Center Program of the
Korea Science and Engineering Foundation through the Center for
Quantum Spacetime (CQUeST) of Sogang University with grant number
R11-2005-021. The work of H.~Shin was supported by grant
No. R01-2004-000-10651-0 from the Basic Research Program of the Korea
Science and Engineering Foundation (KOSEF).



\begin{thebibliography}{99}
\bibitem{fj} R. Floreanini and R. Jackiw, Phys. Rev. Lett. {\bf 59}, 1873 (1987).
\bibitem{sriv} P. P. Srivastava, Phys. Rev. Lett. {\bf 63}, 2791
  (1989); W-T. Kim, J-K. Kim, and Y-J. Park, Phys. Rev. D \textbf{44},
  563 (1991).
\bibitem{bft} I. A. Batalin and E. S. Fradkin, Phys. Lett. B {\bf
  180}, 157 (1986); Nucl. Phys. B {\bf 279}, 514 (1987);
  I. A. Batalin and I. V. Tyutin, Int. J. Mod. Phys. A {\bf 6}, 3255 (1991).
\bibitem{dgml} A.~Das, J.~Gamboa, F.~M{\'e}ndez, and
  J.~L{\'o}pez-Sarri{\'o}n, J.\ High Energy Phys.\ \textbf{05}, 022
  (2004). 
\bibitem{mwy} B. McClain, Y-S. Wu, and F. Yu, Nucl. Phys. B {\bf 343},
  689 (1990).
\bibitem{wot} C. Wotzasek, Phys. Rev. Lett. {\bf 66}, 129 (1991).
\bibitem{an} R.\ Amorim and J.\ Barcelos-Neto, Phys.\ Rev.\ D
  \textbf{53}, 7129 (1996). 
\bibitem{amw} E.\ M.\ C.\ Abreu, R.\ Menezes, and C.\ Wotzasek,
  Phys.\ Rev.\ D \textbf{71}, 065004 (2005). 
\bibitem{ban} R. Banerjee, Phys. Rev. D {\bf 48}, R5467 (1993).
\bibitem{bcs} R. Banerjee, A. Chatterjee, and V. V. Sreedhar,
  Ann. Phys. (N.Y.) {\bf 122}, 254 (1993).
\bibitem{kp} W. T. Kim and Y-J. Park, Phys. Lett. B {\bf 336}, 376 (1994).
\bibitem{brr} R. Banerjee, H. J. Rothe, and K. D. Rothe,
  Phys. Rev. D {\bf 55}, 6339 (1997).
\bibitem{bnon}  R. Banerjee and J. Barcelos-Neto, Nucl. Phys. B {\bf
    499}, 453 (1997); W. Oliveira and J. A. Neto, Nucl. Phys. B {\bf
    533}, 453 (1997).
\bibitem{kkp} W. T. Kim, Y-W. Kim, and Y-J. Park, J. Phys. A {\bf
    32}, 2461 (1999).
\bibitem{kwp} S-K.\ Kim, Y-W.\ Kim, and Y-J.\ Park, Mod.\ Phys.\
 Lett.\ A \textbf{18}, 2287 (2003). 
\bibitem{hkpy} S-T. Hong, W. T. Kim, Y-J. Park, and M. S. Yoon,
  Phys. Rev. D \textbf{62}, 085010 (2000). 
\bibitem{dir} P. A. M. Dirac, {\it Lectures on Quantum Mechanics},
  (Yeshiba University Press, Belfer graduate School, New York, 1964).
\end{thebibliography}
\end{document}